\documentclass[conference]{IEEEtran}
\usepackage{graphicx}
\usepackage{amsmath,amssymb,dsfont,bbm,epstopdf,pgfplots,mathtools,enumitem,mathrsfs, bbm} 
\usepackage{color}
\usepackage{cite}
\usepackage{graphics}
\usepackage{tikz}
\usepackage{pgfplots}
\usepackage{subcaption}
\usetikzlibrary{shapes, arrows, decorations.markings, arrows.meta}
\usetikzlibrary{patterns} 
\allowdisplaybreaks
\graphicspath{{.}{./Figures/}} 
\allowdisplaybreaks[4] 

\newtheorem{corollary}{Corollary}
\newtheorem{theorem}{Theorem}

\newtheorem{remark}{Remark}

\newcommand{\Nak}{{ \mathcal{N}_{\textnormal{active,k}}}}
\newcommand{\Tf}{{\mathcal K_{\text{fast}}} }

\newcommand{\Ta}{{\mathcal K_{\text{active}}} }
\newcommand{\Tfi}{\mathcal{K}_{\textnormal{fast},i}}

\renewcommand{\Pr}{{\mathbb{P}}}

\newcommand{\sRks}{{\{1,\ldots,\lfloor 2^{n R_k^{(S)}} \rfloor \}}}

\renewcommand{\S}{{\sf{S}}}

\renewcommand{\i}{{\iota}}

\newcommand{\MkF}{M_k^{(F)}}
\newcommand{\MkS}{M_k^{(S)}}

\definecolor{ForestGreen}{rgb}{0.0, 0.5, 0.0}

\newcommand{\mw}[1]{{\color{black}#1}}

\newcommand{\hn}[1]{{\color{black}#1}}

\renewcommand{\P}{\mathsf{P}}
\newcommand{\D}{\text{D}}
\renewcommand{\P}{\mathsf{P}}

\newcommand{\sumki}{\S_{\textnormal{sum},i}(\mw{k_{j-1}+1},\ell)} 

\newcommand{\Dt}{\D_{\text {Tx}}}
\newcommand{\Dr}{\D_{\text {Rx}}}

\begin{document}
\title{Cooperative Encoding and Decoding of Mixed Delay Traffic under Random-User Activity}
\author{\IEEEauthorblockN{Homa Nikbakht$^{1}$, Mich\`ele Wigger$^2$, Shlomo Shamai (Shitz)$^3$, Jean-Marie Gorce$^1$}
	\IEEEauthorblockA{$^1$CITI Laboratory, INRIA,   $\quad ^2$LTCI,   T$\acute{\mbox{e}}$l$\acute{\mbox{e}}$com Paris, IP Paris,  $\quad ^3$Technion \\
		\{homa.nikbakht, jean-marie.gorce\}@inria.fr, michele.wigger@telecom-paris.fr,  sshlomo@ee.technion.ac.il}}
\maketitle

 \begin{abstract}
This paper analyses the multiplexing gain (MG) achievable  over Wyner's symmetric network with random user activity and random arrival of mixed-delay traffic. The mixed-delay traffic is composed of delay-tolerant traffic and delay-sensitive traffic where only the former can benefit from  transmitter and receiver cooperation since the latter is subject to stringent decoding delays. The total number of cooperation rounds at transmitter and receiver sides is limited to $\D$ rounds. We derive inner and outer bounds on the MG region. In the limit as $\D\to \infty$, the bounds coincide and the results show that transmitting delay-sensitive messages does not cause any penalty on the sum MG. For finite $\D$ our bounds are still close and prove that the  penalty caused by delay-sensitive transmissions is small.
\end{abstract}

\section{Introduction}
Modern wireless networks have to accommodate a heterogeneous traffic composed of delay-sensitive and delay-tolerant data. For example, communication for remote surgery or other realtime control applications have much more stringent delay  constraints  than communication of standard data. Coding schemes for  such mixed delay traffic are thus of interest to the designers of new generations of wireless networks, notably \cite{shlomo2012ISIT, Zhang2008, Anand2020, Kassab2018, Yin2020,Bairagi, 6Gnew}.  This paper focuses on the mixed-delay multiplexing gain (MG) region of Wyner's symmetric network with randomly activated transmitters (Txs) and receivers (Rxs). The user activity assumption is motivated by random appearance of control or sensor data \mw{or mobility of users}. In our model, Txs and Rxs are allowed to cooperate but only delay-tolerant transmissions can benefit from such cooperation as the cooperation would violate the stringent delay constraints on  delay-sensitive transmissions. \mw{Inherent in this model is the assumption that the cooperation delay dominates the delay introduced by channel coding. Throughout this paper}, we call delay-tolerant messages  ``slow'' messages and delay-sensitive messages  ``fast'' messages. 

Networks with randomly activated users  have been studied previously in \cite{Somekh2008, Levy2008, Somekh2009, HomaITW2020}. Specifically, in our previous work \cite{HomaITW2020}, we  analyzed the MG regions of different interference networks with random user activity and random arrivals of mixed-delay traffic, assuming that only neighbouring receivers can cooperate, but not neighbouring Txs as in this work. Cooperation is assumed to take place over dedicated links and during an unlimited number of rounds. Again,  only  ``slow'' transmissions can benefit from cooperation. The obtained MG regions in \cite{HomaITW2020} showed that  transmitting ``fast'' messages  causes a significant penalty on the sum MG. Notice that an even larger penalty, which grows linearly in the MG of ``fast" messages, applies to any type scheduling algorithm.

 In this paper, we show that this penalty on the sum MG caused by the transmission of ``fast" messages can  be mitigated entirely when not only Rxs but also Txs can cooperate over an unlimited number of rounds. When the number of cooperation rounds is limited to a maximum  number of $\D$ rounds, a small penalty remains, which is however much smaller than when only Rxs can cooperate. Our results in this paper thus show that a joint coding of the two types of messages yields significant benefits in sum-MG as compared to the simpler scheduling algorithms.  
To prove the desired results, we present an information-theoretic converse and propose two coding schemes.  In our first scheme, we schedule  ``fast'' transmissions so that they do not interfere each other. Each  ``fast'' transmission is thus only interfered by ``slow" transmissions, and this interference can  be described to the ``fast" Txs  during the first Tx-cooperation round. This allows the ``fast" Txs to precancel the interference and achieve full MG on each ``fast" Tx.  At the receiver side, ``fast'' Rxs immediately decode their ``fast" messages and  send them  during the first Rx-cooperation round to their neighbours,  which mitigate the interference before decoding their  ``slow'' messages. As a result, ``fast''  messages can be decoded based on interference-free outputs and moreover, they do not disturb the transmission of ``slow''  messages. The transmission of ``slow'' messages can benefit from the remaining  $\D-2$ cooperation rounds, e.g., by applying     \emph{Coordinated Multipoint (CoMP)} reception in small subnets to jointly decode the ``slow'' messages at various receivers. 
Our second scheme only sends  ``slow'' messages. A similar scheme can be used as before, where ``fast" messages can simply be replaced by ``slow" messages. \mw{We obtain an inner bound on the optimal MG region through time-sharing  the two schemes.} 
 

\section{Problem Setup}\label{sec:DescriptionOfTheProblem}



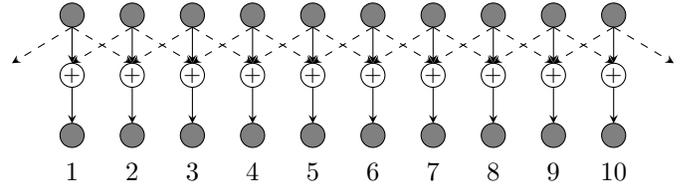
\begin{figure}[t]
  \centering
\begin{tikzpicture}[scale=1.6, >=stealth]
\centering
\tikzstyle{every node}=[draw,shape=circle, node distance=0.5cm];
 \foreach \j in {0,2,4,6,8} {
 \draw [fill= gray](-3.5 + 0.5*\j, 1.5) circle (0.1);
\node[draw =none] (s2) at (-3.5+ 0.5*\j,1 ) {\footnotesize$+$};
\draw (-3.5 +0.5*\j, 1) circle (0.1);
 \draw [fill= gray](-3.5 + 0.5*\j, 0.5) circle (0.1);
 \draw   [->] (-3.5+ 0.5*\j,1.9-0.5)-- (-3.5+ 0.5*\j,1.1);
 \draw   [->] (-3.5+ 0.5*\j,0.9)-- (-3.5+ 0.5*\j,0.6);
  \draw   [->, dashed] (-3.5+ 0.5*\j,1.9-0.5)-- (-3.5+ 0.5*\j + 0.5,1.1);
  \draw   [->, dashed] (-3.5+ 0.5*\j,1.9-0.5)-- (-3.5+0.5*\j - 0.5,1.1);
 }
  \foreach \j in {1,3,5,7,9} {
 \draw [fill= gray](-3.5 + 0.5*\j, 1.5) circle (0.1);
\node[draw =none] (s2) at (-3.5+ 0.5*\j,1 ) {\footnotesize$+$};
\draw (-3.5 +0.5*\j, 1) circle (0.1);
 \draw [fill= gray](-3.5 + 0.5*\j, 0.5) circle (0.1);
 \draw   [->] (-3.5+ 0.5*\j,1.9-0.5)-- (-3.5+ 0.5*\j,1.1);
 \draw   [->] (-3.5+ 0.5*\j,0.9)-- (-3.5+ 0.5*\j,0.6);
  \draw   [->, dashed] (-3.5+ 0.5*\j,1.9-0.5)-- (-3.5+ 0.5*\j + 0.5,1.1);
   \draw   [->, dashed] (-3.5+ 0.5*\j,1.9-0.5)-- (-3.5+ 0.5*\j - 0.5,1.1);
 }
\node[draw =none] (s2) at (-3.5,0.2) {$1$};
\node[draw =none] (s2) at (-3.5+0.5,0.2) {$2$};
\node[draw =none] (s2) at (-3.5+1,0.2) {$3$};
\node[draw =none] (s2) at (-3.5+1.5,0.2) {$4$};
\node[draw =none] (s2) at (-3.5+2,0.2) {$5$};
\node[draw =none] (s2) at (-3.5+2.5,0.2) {$6$};
\node[draw =none] (s2) at (-3.5+3,0.2) {$7$};
\node[draw =none] (s2) at (-3.5+3.5,0.2) {$8$};
\node[draw =none] (s2) at (-3.5+4,0.2) {$9$};
\node[draw =none] (s2) at (-3.5+4.5,0.2) {$10$};
\end{tikzpicture}
  \caption{An illustration of Wyner's symmetric  network with black dashed lines indicating the interference links.} 
   \label{fig2.1b}
   \vspace{-3mm}
  \end{figure}
Consider Wyner's symmetric network with $K$ transmitters (Tx) and ${K}$ receivers (Rx)  that are aligned on two parallel lines so that each Tx $k$  has two neighbours, Tx~$k-1$ and Tx~$k+1$, and each Rx~$k$ has two neighbours, Rx~$k-1$ and Rx~$k+1$. Define $\mathcal K \triangleq \{1, \ldots, K\}$. The signal transmitted by Tx~$k \in \mathcal K$ is observed by Rx~$k$ and the neighboring Rxs $k-1$ and $k+1$.  See Figure~\ref{fig2.1b}.
Each Tx~$k \in \mathcal K$ is \emph{active} with probability $\rho \in [0, 1]$, in which case  it sends a so called ``slow" message $\MkS$ to its corresponding Rx~$k$. Here,  $\MkS$ is uniformly distributed over $\mathcal{M}_{k}^{(S)} \triangleq\sRks$, with $n$ denoting the blocklength and $R_k^{(S)}$ the rate of message $M_k^{(S)}$. Given that Tx~$k$ is active, with probability $\rho_{f}\in [0,1]$,  it  also sends an additional ``fast" message $\MkF$ to Rx~$k$. These ``fast" messages are subject to  stringent delay constraints, as we describe shortly,  
and uniformly distributed over the set $\mathcal{M}^{(F)}\triangleq \{1, \ldots, \lfloor 2^{nR^{(F)}}  \rfloor \}$.  ``Fast" messages are  thus all of same rate $R^{(F)}$. \footnote{\mw{In our model each Tx that has a ``fast" message to send also has a ``slow" message to send. This model is appropriate for systems where ``slow" messages have large volumes and can be delayed over several blocklengths so that most active Txs have ``slow" data to send. The study of a setup where an active Tx might exclusively have ``fast" messages to send is left for future work. See for example \cite{HomaITW2020} for a related work.}}

Let  $A_k=1$ if Tx~$k$ is active and $A_k=0$ if Tx~$k$ is not active. Moreover, if Tx~$k$ is active and has a ``fast" message to send,  set $B_k=1$ and if it is active but has only a ``slow" message to send,  set $B_k=0$. The random tuple $\mathbf{A}:=(A_1, \ldots, A_K)$  is thus independent and identically distributed (i.i.d.) Bernoulli-$\rho$, and if they exist the random variables $B_1, \ldots, B_K$ are  i.i.d Bernoulli-$\rho_f$.  Denote by $\mathbf{B}$ the tuple of $B_k$'s that are defined. Further, define  the \emph{active set} and the \emph{``fast" set} as:
\begin{IEEEeqnarray}{rCl}\label{eq:Ta}
\Ta&  \triangleq& \{k \in  \mathcal K : A_k = 1\},\\
\Tf &\triangleq& \{ k \in \mathcal K\colon  A_k =1 \; \textnormal{and} \; B_k = 1\}.\label{eq:Tf}
\end{IEEEeqnarray}
We  describe the encoding at the active Txs. \mw{The encoding  starts with a first \emph{Tx-cooperation phase} which consists of $\Dt>0$ rounds and depends only on the ``slow" messages in the system. The  ``fast" messages, which are subject to stringent delay constraints, are only  generated afterwards, at the beginning of the subsequent   \emph{channel transmission phase}. So, during the first Tx-cooperation phase, neighbouring active Txs  communicate to each other over dedicated noise-free links of unlimited capacity over $\Dt>0$ rounds. 
 In each cooperation round $j\in\{1,\ldots, \Dt\}$,  any active Tx~$k\in \Ta$ sends  a cooperation message to its active neighbours $\ell \in \mathcal \Nak:=\{k-1,k+1\}\cap \Ta$, where the cooperation message can depend on the Tx's ``slow" message and the cooperation-information  it received during previous rounds. So, in round $j$, Tx~$k$ sends a message}
	\begin{equation}
	T_{k\to \ell}^{(j)}  = \psi_{k\to \ell}^{(j)} \Big(M_{k}^{(S)}, \big\{T_{\ell' \to k}^{(1)}, \ldots, T_{\ell'\to k}^{(j-1)} \big\}_{\ell' \in \Nak},\mathbf{A},\mathbf{B} \Big)
	\end{equation}
	to each Tx~$\ell \in \mathcal \Nak$, for some functions $\psi_{k\to \ell}^{(j)}$ on appropriate domains. 
\mw{At the beginning of the subsequent channel-coding phase, ``fast" messages are generated and each active} $k \in \mathcal K$, Tx~$k$ computes its channel inputs $ X_k^n\triangleq ( X_{k,1},\ldots, X_{k,n}) \in \mathbb R^{n}$ as 
\begin{IEEEeqnarray}{rCl}
X_{k}^n = \begin{cases}f_k^{(B)} \big( \MkF, \MkS,  \{T_{\ell' \to k}^{(j)}\},\mathbf{A},\mathbf{B}\big), & k \in \Tf \\
f_k^{(S)} \big( \MkS,  \{T_{\ell' \to k}^{(j)}\}, \mathbf{A},\mathbf{B}\big), & \hspace{-.7cm} k \in(\Ta \backslash \Tf)\\
0, &  \hspace{-.3cm}  k \in (\mathcal{K}\backslash \Ta).
\end{cases} \nonumber\\
\end{IEEEeqnarray}
\mw{where the sets are over  $j \in \{1, \ldots, \Dt \}$ and  $\ell' \in \Nak$, and  $f_k^{(B)}$ and  $f_k^{(S)}$ are  encoding functions  on appropriate domains satisfying }
the average block-power constraint
\begin{equation}\label{eq:power}
\frac{1}{n} \sum_{t=1}^n X_{k,t}^2
\leq \P, 
\quad \forall\ k \in \mathcal K, \qquad \textnormal{almost surely.}
\end{equation}

The input-output relation of the network is  described as
\begin{equation}\label{Eqn:Channel}
{Y}_{k,t} = A_k  {X}_{k,t} +\sum_{\tilde k \in \{k-1, k+1\}} A_{\tilde k} h_{\tilde k , k} {X}_{ \tilde k,t}+ {Z}_{k,t},
\end{equation}
where $\{Z_{k,t}\}$ are independent and identically distributed (i.i.d.) standard Gaussians for all $k$ and $t$ and independent of all messages; $h_{\tilde k , k}>0$ with $\tilde k \in \{k-1, k+1\}$ is the channel coefficient between Tx~$\tilde k$ and Rx~$k$ and is  a fixed real number smaller than $1$; and $X_{0,t} = 0$ for all $t$.

\mw{Decoding also takes place in two phases. In the first \emph{``fast"-decoding phase}, any active  Rx~$k \in \Tf$ decodes the ``fast" message $M_k^{(F)}$ based on its own channel outputs $Y_k^n$ by computing:}
\begin{align}
	&\hat{{{M}}} _k^{(F)} ={g_k^{(n)}}\big( Y_k^{n}\big),
	\end{align} 
for some decoding function $g_k^{(n)}$ on appropriate domains.	
In the subsequent \emph{slow-decoding phase},  active Rxs first communicate with their active neighbours  during $\Dr \ge 0$ rounds over dedicated noise-free links with unlimited capacity, and then they decode their intended ``slow" messages based on their outputs and based on this exchanged information.   
Specifically, in each cooperation round $j\in \{1, \ldots, \Dr \}$,  each active Rx~$k\in \mathcal T_{\text{active}}$ sends  a cooperation message  
\mw{\begin{equation}
Q^{(j)}_{k\rightarrow \ell}=\phi_{k\to \ell}^{(j)}\Big( \mathbf Y_k^n,\big\{ Q^{(1)}_{\ell' \rightarrow k},\ldots , Q^{(j-1)}_{\ell' \rightarrow k}\}_{\ell' \in \Nak},\mathbf{A},\mathbf{B}\Big)
\end{equation}
to Rx~$\ell$ if  $\ell\in  \Nak$ for some appropriate function $\phi_{k\to \ell}^{(j)}$.}

After the last cooperation round, each active Rx~$k\in \Ta$ decodes its desired ``slow" messages as
\begin{equation}\label{mhats}
\hat{{M}}_{k}^{(S)}={b_{k}^{(n)}}\Big( \mathbf Y_{k}^n, \Big\{ Q^{(1)}_{\ell'\rightarrow k}, \ldots, Q^{(\Dr)}_{\ell'\rightarrow k}\Big\}_{\ell' \in \Nak},\mathbf{A},\mathbf{B}\Big),
\end{equation}
 where  $b_{k}^{(n)}$ is a decoding function on appropriate domains. 

The maximum number of Tx-cooperation rounds $\Dt$ and Rx-cooperation rounds $\Dr$ are design parameters but subject to a total delay constraint:
\begin{equation}\label{eq:Dsum}
\Dt +\Dr \leq \D, 
\end{equation}
	for a given $\D \geq 0$. 
		
Given $\P>0$ and $K>0$, a rate pair $(R^{(F)}(\P), \bar{R}_K^{(S)}(\P))$ is said $\D$-achievable if there exist rates $\{R_{k}^{(S)}\}_{k=1}^{K}$ satisfying
\begin{IEEEeqnarray}{rCl}
\bar R_K^{(S)} &\le &  \mathbb E \left [\sum_{k \in \Ta } R_{k}^{(S)} \right ] ,
\end{IEEEeqnarray} 
a pair of Tx- and Rx-cooperation rounds $\Dt,\Dr$ summing to $\Dt+\Dr=\D$ and
encoding, cooperation, and decoding functions   satisfying constraint \eqref{eq:power} and so that the probability of error
\begin{equation}
\Pr \bigg[\bigcup_{k \in \Tf} \!\!\big( \hat{M}_k^{(F)} \neq M_k^{(F)}\big)  \;\text{or} \! \!\bigcup_{k \in \Ta} \big(\hat{M}_k^{(S)} \neq M_k^{(S)}\big)  \bigg]  
\end{equation}
tends to $0$ as $n\to \infty$.
An MG pair  $(\S^{(F)},\S^{(S)})$ is called $\D$-\emph{achievable}, if for all powers $\P>0$ there exist $\D$-achievable  rates   $\big\{{R}_K^{(F)}(\P),\bar R_K^{(S)}(\P) \big\}_{\P>0}$
satisfying
\begin{IEEEeqnarray} {rCl}
\S^{(F)}& \triangleq&\varlimsup_{K\rightarrow \infty}\varlimsup_{\P\rightarrow\infty} \;\frac{ R_K^{(F)}(\P)}{\frac{K}{2}\log (\P)} \cdot \rho \rho_f ,\label{eq:sf}\\
\S^{(S)}& \triangleq&\varlimsup_{K\rightarrow \infty}  \varlimsup_{\P\rightarrow\infty} \; \frac{  \bar R_K^{(S)}(\P)}{\frac{K}{2}\log (\P)} .\label{eq:ss}
\end{IEEEeqnarray}
The  closure of the set of all achievable   MG pairs $(\S^{(F)}, \S^{(S)})$ is called $\D$-\emph{cooperative fundamental MG region} and is denoted $\mathcal{S}_{\D}^\star(\rho, \rho_f)$. 

The MG in \eqref{eq:ss} measures the average expected ``slow'' MG on the network. Since the ``fast" rate is fixed to $R^{(F)}$ at all Txs  in $ \Tf$, we multiply the MG in \eqref{eq:sf} by $\rho  \rho_f$ to obtain the average expected ``fast" MG of the network.  

\section{Main Results}
Our first result is an inner bound on  $\mathcal{S}_{\D}^\star(\rho, \rho_f)$. It is based on \mw{time-sharing} two schemes, one with large ``fast'' MG  and the other with zero ``fast'' MG.

  \begin{theorem}[Inner Bound on MG Region] \label{theorem1}
 For $\rho\in (0,1)$, the  fundamental MG region $\S_{\D}^*(\rho, \rho_f)$ includes all nonnegative pairs $(\S^{(F)}, \; \S^{(S)})$ satisfying
 \begin{IEEEeqnarray}{rCl}
 \S^{(F)} &\le& \frac{\rho \rho_f}{2}, \label{eq:a} \\
 \S^{(S)} + M \cdot  \S^{(F)} & \le & \rho - \frac{(1-\rho) \rho^{\D+2}}{1-\rho^{\D+2}},
 \end{IEEEeqnarray}
 where 
 \begin{IEEEeqnarray}{rCl} \label{eq:M}
 M \triangleq 1+ \frac{(1-\rho)^2 \rho^{\D+2}}{\rho \rho_f(1-\rho^{\D+2})} +\frac{(1-\rho)^2  \rho^{\D+1}  (1-\rho_f)^{\frac{\D}{2}}}{\rho \rho_f (1- \rho^{\D+2}(1-\rho_f)^{\frac{\D}{2}+1})}. \IEEEeqnarraynumspace
 \end{IEEEeqnarray}
For $\rho=1$, it includes all pairs satisfying \eqref{eq:a} and 
 \begin{IEEEeqnarray}{rCl}\label{eq:rho1}
 \S^{(S)} +\S^{(F)} & \le &\frac{\D+1}{\D+2}. \label{eq:rhoequal1}
 \end{IEEEeqnarray}
 \end{theorem}
 \begin{IEEEproof}
 See Section \ref{sec:achiv}.
 \end{IEEEproof} 
 We also have the following outer bound.
 \begin{theorem}[Outer Bound on MG Region] \label{theorem2}
 For $\rho \in (0,1)$, all achievable MG pairs  $(\S^{(F)}, \; \S^{(S)})$ satisfy \eqref{eq:a} and 
 \begin{IEEEeqnarray}{rCl}
 \S^{(S)} + \S^{(F)} & \le & \rho - \frac{(1-\rho) \rho^{\D+2}}{1-\rho^{\D+2}}. \label{eq:conv2}
 \end{IEEEeqnarray}
 For $\rho =1$ they satisfy \eqref{eq:a} and \eqref{eq:rho1}.
 \end{theorem}
 \begin{IEEEproof}
See Section \ref{sec:converse}. \end{IEEEproof}
 
 Inner and outer bounds are generally very close. They  coincide in the extreme cases $\rho=1$ and $\D \to \infty$.
 \begin{corollary}
For $\rho = 1$ or when $\D\to \infty$, Theorem~\ref{theorem2} is exact. For $\rho=1$, 
the fundamental MG region  $\S^*(\rho, \rho_f)$ \emph{is} the set of  all nonnegative MG pairs $(\S^{(F)}, \; \S^{(S)})$ satisfying \eqref{eq:a} and \eqref{eq:rho1}, and for $\rho\in(0,1)$ \mw{and $\D\to \infty$} it \emph{is} the set of all MG pairs  $(\S^{(F)}, \; \S^{(S)})$ satisfying \eqref{eq:a} and 
 \begin{IEEEeqnarray}{rCl}
 \S^{(S)} +\S^{(F)} & \le & \rho.
 \end{IEEEeqnarray}
\end{corollary}
 
 \begin{remark}
 In our model, we assume that  neighbouring Txs and neighbouring Rxs can only cooperate if they lie in the active set $\Ta$. Txs and Rxs in the \emph{inactive set} $\mathcal{K}\backslash \Ta$ do not participate in the cooperation phases. Notice that all our results remain valid in a setup where inactive Txs and Rxs \emph{do} participate in the cooperation phases. Since our inner and outer bounds are rather close in general (see the subsequent numerical discussion), this indicates that without essential loss in optimality Txs and Rxs in $\mathcal{K}\backslash \Ta$  can entirely be set to sleep mode to conserve their batteries.
 \end{remark}

 \begin{figure}[t!]
\centering
\begin{tikzpicture}[scale=.9]
\begin{axis}[
    xlabel={\small {$\S^{(F)}$ }},
    ylabel={\small {$\S^{(S)}$ }},
     xlabel style={yshift=.5em},
     ylabel style={yshift=-1.25em},
    xmin=0, xmax=0.27,
    ymin=0, ymax=0.82,
    xtick={0,0.1,0.2,0.3,0.4},
    ytick={0,0.1,0.2,0.3,0.4,0.5,0.6,0.7,0.8,0.9,1},
    yticklabel style = {font=\small,xshift=0.25ex},
    xticklabel style = {font=\small,yshift=0.25ex},
    legend pos=south west,
]

 \addplot[ color=green,   mark=star, line width = 0.5mm] coordinates {  (0,0.8) (0.24, 0.56)(0.24,0) };
 \addplot[ color=black,   mark=star, line width = 0.5mm, dashed] coordinates {  (0,0.7852) (0.24, 0.5452)(0.24,0) };
      \addplot[ color=blue,   mark=halfcircle, thick] coordinates {  (0,0.7852) (0.24, 0.5434)(0.24,0) };
   \addplot[ color=brown,   mark=diamond,  line width = 0.5mm, dashed] coordinates {  (0,0.7289) (0.24, 0.4889)(0.24,0) };
     \addplot[ color=red,   mark=halfcircle, thick] coordinates {  (0,0.7289) (0.24, 0.4790)(0.24,0) };


\small 
      \legend{{Outer and Inner Bounds, $\D = \infty$}, {Outer Bound, $\D = 10, M = 1$}, {Inner Bound, $ \D = 10, M = 1.006$},{Outer Bound, $\D = 4, M = 1$},  {Inner Bound, $\D = 4, M = 1.03$}}  
\end{axis}


\vspace{-0.4cm}
\end{tikzpicture}

\caption{Inner and outer bounds on  MG region $\mathcal{S}_{\D}^\star(\rho, \rho_f)$ for $\rho= 0.8$ and $\rho_f = 0.6$, and different values of $\D$.}
\label{fig2itw}
\vspace{-0.5cm}
\end{figure}
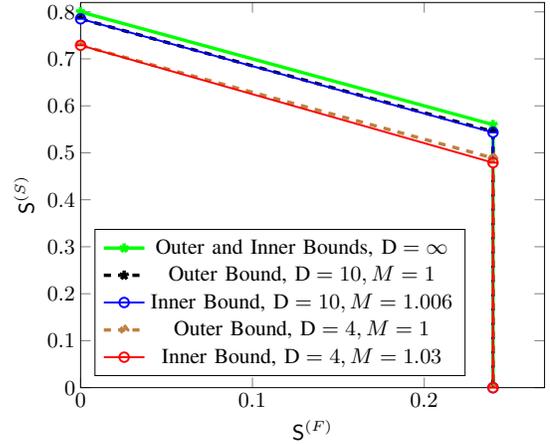

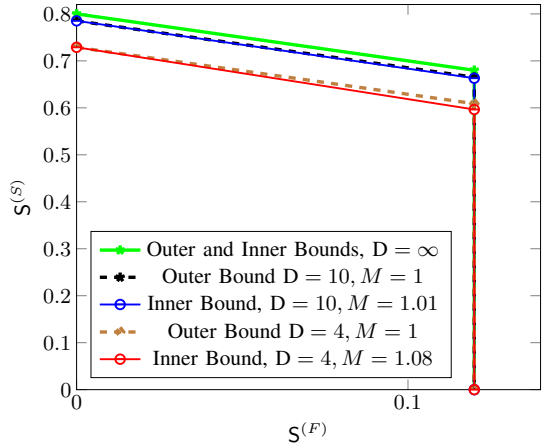
\begin{figure}[t!]
\centering
\begin{tikzpicture}[scale=.9]
\begin{axis}[
    xlabel={\small {$\S^{(F)}$ }},
    ylabel={\small {$\S^{(S)}$ }},
     xlabel style={yshift=.5em},
     ylabel style={yshift=-1.25em},
    xmin=0, xmax=0.14,
    ymin=0, ymax=0.82,
    xtick={0,0.1,0.2,0.3,0.4},
    ytick={0,0.1,0.2,0.3,0.4,0.5,0.6,0.7,0.8,0.9,1},
    yticklabel style = {font=\small,xshift=0.25ex},
    xticklabel style = {font=\small,yshift=0.25ex},
    legend pos=south west,
]

 \addplot[ color=green,   mark=star, line width = 0.5mm] coordinates {  (0,0.8) (0.12, 0.68)(0.12,0) };
 \addplot[ color=black,   mark=star, line width = 0.5mm, dashed] coordinates {  (0,0.7852) (0.12, 0.6652)(0.12,0) };
      \addplot[ color=blue,   mark=halfcircle, thick] coordinates {  (0,0.7852) (0.12, 0.6631)(0.12,0) };

 \addplot[ color=brown,   mark=diamond,  line width = 0.5mm, dashed] coordinates {  (0,0.7289) (0.12, 0.6089)(0.12,0) };
     \addplot[ color=red,   mark=halfcircle, thick] coordinates {  (0,0.7289) (0.12, 0.5965)(0.12,0) };

\small 
      \legend{{Outer and Inner Bounds, $\D = \infty$}, {Outer Bound $\D = 10, M = 1$},{Inner Bound, $\D = 10, M = 1.01$}, {Outer Bound $\D = 4, M = 1$}, {Inner Bound, $\D = 4, M = 1.08$}}  
\end{axis}


\vspace{-0.4cm}
\end{tikzpicture}

\caption{Inner and outer bounds on MG Region $\mathcal{S}_{\D}^\star(\rho, \rho_f)$ for $\rho= 0.8$, $\rho_f = 0.3$  and different values of $\D$.}
\label{fig3itw}

\end{figure}

Figures \ref{fig2itw}--\ref{fig4itw} illustrate the outer and inner bounds on the MG region for different values of $\rho, \rho_f$, and $\D$. The bounds all have 
maximum ``fast" MG $\S^{(F)}=\frac{\rho \rho_f}{2}$.  Obviously, all {bounds 
 increase} with the activity parameter $\rho$. The  most interesting part of the bounds is the upper side of the {trapezoids}, which lies opposite the two right angles. In particular, the slope of this {side}, which is $-1$ for the outer bounds and $-M$ for the inner bounds,  describes the penalty in sum MG $\S^{(F)}+\S^{(S)}$ incurred when one  increases the ``fast" MG $\S^{(F)}$. In the outer bounds, the sum MG along this line stays constant for all values  of the ``fast'' MG $\S^{(F)}$. In our inner bounds, the sum-MG is reduced by $(M-1) \S$ when the ``fast" MG is increased by $\S$.  This penalty decreases  as $\D$ increases, and is already negligible for $\D=10$ as the three figures illustrate. In fact, for $\D=10$ the MG region achieved by our inner bounds is close to the limiting MG regions for $\D \to \infty$, indicating that increasing the number of cooperation rounds beyond 10 provides only a marginal gain in MG region. As seen in Figure~\ref{fig4itw}, for small user activity parameter $\rho$ even a small number of cooperation rounds  ($\D=4$) suffices to well approximate the asymptotic MG region for $\D\to \infty$.  The reason is that a large number of cooperation rounds is only useful in subnets with a large number of consecutive Txs that are  active, and such subnets are extraordinarily rare when $\rho$ is small. 
Figures~\ref{fig2itw} and \ref{fig3itw} further indicate that the penalty in maximum sum-MG of our inner bounds also decreases when the ``fast" activity parameter $\rho_f$ increases. For example,  for $\rho=0.6$ and $\D = 4$  the sum-MG penalty $(M-1)$ of the inner bound decreases from $0.08$ for $\rho_f=0.3$ to $0.03$ for $\rho_f=0.6$ (see Figures \ref{fig3itw} and \ref{fig2itw}).

In our previous work  \cite[Theorem~2]{HomaITW2020} we studied the MG region of the present network but with only Rx-conferencing. In contrast to our results here, in \cite{HomaITW2020} there is always a penalty on the sum-MG when  transmitting at positive ``fast'' MGs. These results indicate that the sum-MG penalty caused by the ``fast" transmssions can  only be mitigated when both  Txs and Rxs can cooperate, but Rx cooperation alone is not sufficient. In fact, in our schemes we mitigate interference from ``fast" transmissions on ``slow" transmissions via Rx-cooperation and we mitigate interference from ``slow" transmissions on ``fast" transmissions via Tx-cooperation.  In \cite{HomaITW2020} we could only mitigate the former interference but not the latter. 

 \begin{figure}[t!]
\centering
\begin{tikzpicture}[scale=.9]
\begin{axis}[
    xlabel={\small {$\S^{(F)}$ }},
    ylabel={\small {$\S^{(S)}$ }},
     xlabel style={yshift=.5em},
     ylabel style={yshift=-1.25em},
    xmin=0, xmax=0.065,
    ymin=0, ymax=0.42,
    xtick={0,0.05, 0.1,0.2,0.3,0.4},
    ytick={0,0.1,0.2,0.3,0.4,0.5,0.6,0.7,0.8,0.9,1},
    yticklabel style = {font=\small,xshift=0.25ex},
    xticklabel style = {font=\small,yshift=0.25ex},
      legend pos=south west,
]

 \addplot[ color=green,   mark=star, line width = 0.85mm] coordinates {  (0,0.4) (0.06, 0.34)(0.06,0) };
 \addplot[ color=black,   mark=star, line width = 0.5mm, dashed] coordinates {  (0,0.4) (0.06, 0.34)(0.06,0) };

      \addplot[ color=blue,   mark=halfcircle, thick] coordinates {  (0,0.4) (0.06, 0.34)(0.06,0) };
       \addplot[ color=brown,   mark=diamond,  line width = 0.5mm, dashed] coordinates {  (0,0.3975) (0.06,0.3375)(0.06,0) };

     \addplot[ color=red,   mark=halfcircle, thick] coordinates {  (0,0.3975) (0.06,0.3348)(0.06,0) };
     


\small 
      \legend{{Outer and Inner Bounds, $\D = \infty$},{Outer Bound, $\D = 10, M = 1$},{Inner Bound, $\D = 10, M = 1.0001$},{Outer Bound, $\D = 4, M = 1$},  {Inner Bound, $\D = 4, M = 1.027$}}  
\end{axis}

\end{tikzpicture}

\caption{MG Region $\mathcal{S}_{\D}^\star(\rho, \rho_f)$ for  $\rho = 0.4$, $\rho_f= 0.3$ and different values of  $\D$.}
\label{fig4itw}
\vspace{-3mm}

\end{figure}
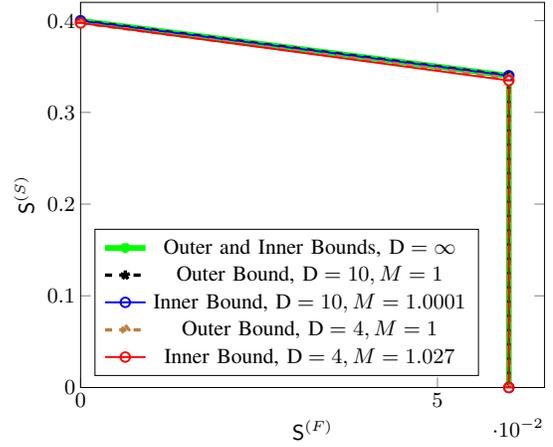

\begin{figure*}[t]
  \centering
\begin{subfigure}{1\textwidth}
\centering
\begin{tikzpicture}[scale=1.6, >=stealth]
\centering
\tikzstyle{every node}=[draw,shape=circle, node distance=0.5cm];
 \foreach \j in {1,...,20} {
 \draw (-3.5 + 0.5*\j, 1.5) circle (0.1);
\node[draw =none] (s2) at (-3.5+ 0.5*\j,1 ) {\footnotesize$+$};
\draw (-3.5 +0.5*\j, 1) circle (0.1);
 \draw (-3.5 + 0.5*\j, 0.5) circle (0.1);
 \draw   [->] (-3.5+ 0.5*\j,1.9-0.5)-- (-3.5+ 0.5*\j,1.1);
 \draw   [->] (-3.5+ 0.5*\j,0.9)-- (-3.5+ 0.5*\j,0.6);
 }
  \foreach \j in {1,...,6,10,14,15,16,17,18} {
  \draw   [->, dashed] (-3.5+ 0.5*\j,1.9-0.5)-- (-3.5+ 0.5*\j + 0.5,1.1);
  }
    \foreach \j in {2,...,7,11,15,16,17,18,19} {
  \draw   [->, dashed] (-3.5+ 0.5*\j,1.9-0.5)-- (-3.5+0.5*\j - 0.5,1.1);
 }
  \foreach \j in {1,3,7,11,15} {
 \draw [fill = red](-3.5 + 0.5*\j, 1.5) circle (0.1);
\node[draw =none] (s2) at (-3.5+ 0.5*\j,1 ) {\footnotesize$+$};
\draw (-3.5 +0.5*\j, 1) circle (0.1);
 \draw [fill = red] (-3.5 + 0.5*\j, 0.5) circle (0.1);
 \draw   [->] (-3.5+ 0.5*\j,1.9-0.5)-- (-3.5+ 0.5*\j,1.1);
 \draw   [->] (-3.5+ 0.5*\j,0.9)-- (-3.5+ 0.5*\j,0.6);
 }
   \foreach \j in {2,4,5,6,10,14,16,17, 18,19} {
 \draw [fill = blue](-3.5 + 0.5*\j, 1.5) circle (0.1);
\node[draw =none] (s2) at (-3.5+ 0.5*\j,1 ) {\footnotesize$+$};
\draw (-3.5 +0.5*\j, 1) circle (0.1);
 \draw [fill = blue] (-3.5 + 0.5*\j, 0.5) circle (0.1);
 \draw   [->] (-3.5+ 0.5*\j,1.9-0.5)-- (-3.5+ 0.5*\j,1.1);
 \draw   [->] (-3.5+ 0.5*\j,0.9)-- (-3.5+ 0.5*\j,0.6);
 }
 \node[draw =none] (s2) at (-3.5+0.5,0.2) {$1$};
\node[draw =none] (s2) at (-3.5+1,0.2) {$2$};
\node[draw =none] (s2) at (-3.5+1.5,0.2) {$3$};
\node[draw =none] (s2) at (-3.5+2,0.2) {$4$};
\node[draw =none] (s2) at (-3.5+2.5,0.2) {$5$};
\node[draw =none] (s2) at (-3.5+3,0.2) {$6$};
\node[draw =none] (s2) at (-3.5+3.5,0.2) {$7$};
\node[draw =none] (s2) at (-3.5+4,0.2) {$8$};
\node[draw =none] (s2) at (-3.5+4.5,0.2) {$9$};
\node[draw =none] (s2) at (-3.5+5,0.2) {$10$};
\node[draw =none] (s2) at (-3.5+5.5,0.2) {$11$};
\node[draw =none] (s2) at (-3.5+6,0.2) {$12$};
\node[draw =none] (s2) at (-3.5+6.5,0.2) {$13$};
\node[draw =none] (s2) at (-3.5+7,0.2) {$14$};
\node[draw =none] (s2) at (-3.5+7.5,0.2) {$15$};
\node[draw =none] (s2) at (-3.5+8,0.2) {$16$};
\node[draw =none] (s2) at (-3.5+8.5,0.2) {$17$};
\node[draw =none] (s2) at (-3.5+9,0.2) {$18$};
\node[draw =none] (s2) at (-3.5+9.5,0.2) {$19$};
\node[draw =none] (s2) at (-3.5+10,0.2) {$20$};

\node[draw =none] (s2) at (-3.5+0.5,1.75) {$F$};
\node[draw =none] (s2) at (-3.5+1,0.2+1.55) {$S$};
\node[draw =none] (s2) at (-3.5+1.5,0.2+1.55) {$F$};
\node[draw =none] (s2) at (-3.5+2,0.2+1.55) {$S$};
\node[draw =none] (s2) at (-3.5+2.5,0.2+1.55) {$S$};
\node[draw =none] (s2) at (-3.5+3,0.2+1.55) {$S$};
\node[draw =none] (s2) at (-3.5+3.5,0.2+1.55) {$F$};
\node[draw =none] (s2) at (-3.5+5,0.2+1.55) {$S$};
\node[draw =none] (s2) at (-3.5+5.5,0.2+1.55) {$F$};
\node[draw =none] (s2) at (-3.5+7,0.2+1.55) {$S$};
\node[draw =none] (s2) at (-3.5+7.5,0.2+1.55) {$F$};
\node[draw =none] (s2) at (-3.5+8,0.2+1.55) {$S$};
\node[draw =none] (s2) at (-3.5+8.5,0.2+1.55) {$S$};
\node[draw =none] (s2) at (-3.5+9,0.2+1.55) {$S$};
\node[draw =none] (s2) at (-3.5+9.5,0.2+1.55) {$S$};

 \node[draw =none, rotate =45] (s2) at ( 0.5,1.5) {\huge$+$};
  \node[draw =none, rotate =45] (s2) at ( 0.5,0.5) {\huge$+$};
  \node[draw =none, rotate =45] (s2) at ( 6.5,1.5) {\huge$+$};
  \node[draw =none, rotate =45] (s2) at (6.5,0.5) {\huge$+$};
  \foreach \i in {1,3,15} {
  \draw [<-, very thick, blue ] (-3.5 + 0.5*\i+0.1, 1.5) -- (-3.5 + 0.5*\i + 0.5- 0.1, 1.5);
  }
   \foreach \i in {2,6,10,14,18} {
  \draw [->, very thick, blue ] (-3.5 + 0.5*\i+0.1, 1.5) -- (-3.5 + 0.5*\i + 0.5- 0.1, 1.5);
  }
  
   \foreach \i in {2,6,10,14} {
  \draw [<-, very thick, red ] (-3.5 + 0.5*\i+0.1, 0.5) -- (-3.5 + 0.5*\i + 0.5- 0.1, 0.5);
  }
  
    \foreach \i in {1,15} {
  \draw [->, very thick, red ] (-3.5 + 0.5*\i+0.1, 0.5) -- (-3.5 + 0.5*\i + 0.5- 0.1, 0.5);
  }
     \foreach \i in {18} {
  \draw [<-, very thick, blue ] (-3.5 + 0.5*\i+0.1, 0.5) -- (-3.5 + 0.5*\i + 0.5- 0.1, 0.5);
  }
   \foreach \j in {4,16} {
 \draw [very thick, green](-3.5 + 0.5*\j, 0.5) circle (0.12);
  }
\end{tikzpicture}
\caption{The first Tx-cooperation round and the first Rx-cooperation round}
\end{subfigure}

\begin{subfigure}{1\textwidth}
\centering
\begin{tikzpicture}[scale=1.6, >=stealth]
\centering
\tikzstyle{every node}=[draw,shape=circle, node distance=0.5cm];
 \foreach \j in {2,4,5,6,14,16,17,18} {
 \draw (-3.5 + 0.5*\j, 1.5) circle (0.1);
\node[draw =none] (s2) at (-3.5+ 0.5*\j,1 ) {\footnotesize$+$};
\draw (-3.5 +0.5*\j, 1) circle (0.1);
 \draw (-3.5 + 0.5*\j, 0.5) circle (0.1);
 \draw   [->] (-3.5+ 0.5*\j,1.9-0.5)-- (-3.5+ 0.5*\j,1.1);
 \draw   [->] (-3.5+ 0.5*\j,0.9)-- (-3.5+ 0.5*\j,0.6);
 }
  \foreach \j in {4,5,16,17} {
  \draw   [->, dashed] (-3.5+ 0.5*\j,1.9-0.5)-- (-3.5+ 0.5*\j + 0.5,1.1);
  }
   \foreach \j in {2,14} {
  \draw   [->, dashed] (-3.5+ 0.5*\j,1.9-0.5)-- (-3.5+ 0.5*\j + 0.5+0.5,1.1);
  }
    \foreach \j in {5,6,17,18} {
  \draw   [->, dashed] (-3.5+ 0.5*\j,1.9-0.5)-- (-3.5+0.5*\j - 0.5,1.1);
 }
   \foreach \j in {4,16} {
  \draw   [->, dashed] (-3.5+ 0.5*\j,1.9-0.5)-- (-3.5+0.5*\j - 0.5-0.5,1.1);
 }
  \foreach \j in {1,3,7,11,15,19} {
 \draw [fill = white, draw = none](-3.5 + 0.5*\j, 1.5) circle (0.1);
 }
   \foreach \j in {2,4,5,6,14,16,17, 18} {
 \draw [fill = blue](-3.5 + 0.5*\j, 1.5) circle (0.1);
\node[draw =none] (s2) at (-3.5+ 0.5*\j,1 ) {\footnotesize$+$};
\draw (-3.5 +0.5*\j, 1) circle (0.1);
 \draw [fill = blue] (-3.5 + 0.5*\j, 0.5) circle (0.1);
 \draw   [->] (-3.5+ 0.5*\j,1.9-0.5)-- (-3.5+ 0.5*\j,1.1);
 \draw   [->] (-3.5+ 0.5*\j,0.9)-- (-3.5+ 0.5*\j,0.6);
 }
\node[draw =none] (s2) at (-3.5+1,0.2) {$2$};
\node[draw =none] (s2) at (-3.5+2,0.2) {$4$};
\node[draw =none] (s2) at (-3.5+2.5,0.2) {$5$};
\node[draw =none] (s2) at (-3.5+3,0.2) {$6$};
\node[draw =none] (s2) at (-3.5+7,0.2) {$14$};
\node[draw =none] (s2) at (-3.5+8,0.2) {$16$};
\node[draw =none] (s2) at (-3.5+8.5,0.2) {$17$};
\node[draw =none] (s2) at (-3.5+9,0.2) {$18$};
\node[draw =none] (s2) at (-3.5+1,0.2+1.55) {$S$};
\node[draw =none] (s2) at (-3.5+2,0.2+1.55) {$S$};
\node[draw =none] (s2) at (-3.5+2.5,0.2+1.55) {$S$};
\node[draw =none] (s2) at (-3.5+3,0.2+1.55) {$S$};
\node[draw =none] (s2) at (-3.5+7,0.2+1.55) {$S$};
\node[draw =none] (s2) at (-3.5+8,0.2+1.55) {$S$};
\node[draw =none] (s2) at (-3.5+8.5,0.2+1.55) {$S$};
\node[draw =none] (s2) at (-3.5+9,0.2+1.55) {$S$};

 \node[draw =none, rotate =45] (s2) at ( 0.5,1.5) {\huge${\color{white}+}$};
  \node[draw =none, rotate =45] (s2) at ( 0.5,0.5) {\huge${\color{white}+}$};
  \node[draw =none, rotate =45] (s2) at ( 6.5,1.5) {\huge${\color{white}+}$};
  \node[draw =none, rotate =45] (s2) at (6.5,0.5) {\huge${\color{white}+}$};
%
%
     \foreach \i in {4,5,16,17} {
  \draw [<-, very thick, blue ] (-3.5 + 0.5*\i+0.1, 0.5-0.05) -- (-3.5 + 0.5*\i + 0.5- 0.1, 0.5-0.05);
   \draw [->, very thick, blue ] (-3.5 + 0.5*\i+0.1, 0.5+0.05) -- (-3.5 + 0.5*\i + 0.5- 0.1, 0.5+0.05);
  }
   \foreach \i in {2,14} {
  \draw [<-, very thick, blue ] (-3.5 + 0.5*\i+0.1, 0.5-0.05) -- (-3.5 + 0.5*\i + 0.5, 0.5-0.05);
   \draw [<-, very thick, blue ] (-3.5 + 0.5*\i+0.5, 0.5-0.05) -- (-3.5 + 0.5*\i + 0.5+0.5, 0.5-0.05);
   \draw [->, very thick, blue ] (-3.5 + 0.5*\i, 0.5+0.05) -- (-3.5 + 0.5*\i + 0.5, 0.5+0.05);
    \draw [->, very thick, blue ] (-3.5 + 0.5*\i+0.5, 0.5+0.05) -- (-3.5 + 0.5*\i + 0.5+0.5-0.1, 0.5+0.05);
  }
   \foreach \j in {4,16} {
 \draw [very thick, green](-3.5 + 0.5*\j, 0.5) circle (0.12);
  }
\end{tikzpicture}
\vspace{-0.3cm}
\caption{The last $\D-2$ Rx-cooperation rounds.} 
\end{subfigure}

  \caption{Example for $\D = 6$ : Tx/Rx pairs in red  have ``fast'' messages to transmit, Tx/Rx pairs in blue have ``slow''messages to transmit, Tx/Rx pairs in white are deactivated. We deactivated Txs/Rxs pairs $8$ and $20$ to satisfy the delay constraint $\D$. Rx~$4$ and Rx~$16$ are master Rxs. Tx/Rx pair $19$ employs the same coding scheme as the ``fast'' transmissions.}
   \label{fig-exp1}
   \vspace{-3mm}
  \end{figure*}
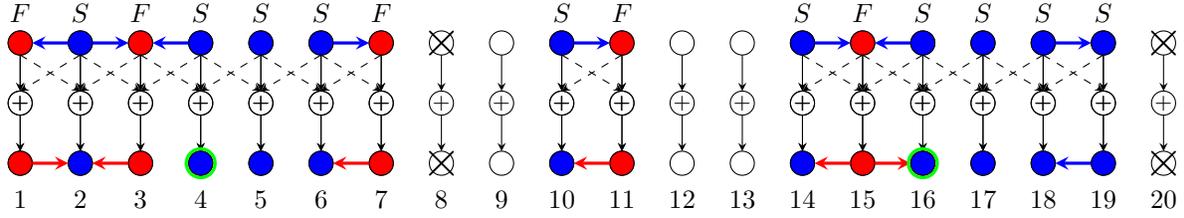
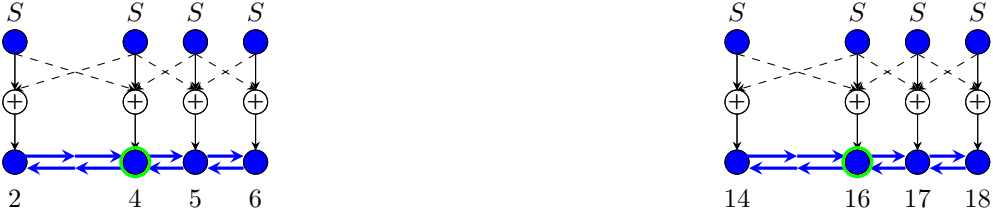
\section{Proof of Achievability of Theorem \ref{theorem1}} \label{sec:achiv}
We describe two schemes, which through time-sharing arguments establish the achievability of the inner bound in Theorem~\ref{theorem1}. The first scheme transmits at maximum $\S^{(F)}=\frac{\rho \rho_f}{2}$, and the second scheme at $\S^{(F)}=0$.
Both schemes divide the maximum number of  cooperation rounds $\D$ into Tx-cooperation and Rx-cooperation rounds as:  
\begin{equation}
\Dt=1 \quad \textnormal{and} \quad \Dr= \D-1.
\end{equation}

 For simplicity we assume $\D$ and $K$ even. 
 
 \subsection{Scheme 1: Transmitting at large $\S^{(F)}$} \label{schemef}

 We partition $\mathcal K$  into $2$ groups $\mathcal K_1$ and $ \mathcal K_{2}$,
 \begin{IEEEeqnarray}{rCl}
 	\mathcal K_1 &\triangleq& \{1,3, \ldots, K-1\}, \\
 	\mathcal K_2 &\triangleq& \{2,4, \ldots, K\},
 \end{IEEEeqnarray}
 so that  all the signals sent by  Txs in a group  $\mathcal{K}_i$ do not interfere with each other, for  $i = 1,2$. 
 We further divide the total \mw{channel} transmission time into two equally-sized phases. \mw{(These phases can be interleaved or subsequent, and both take place after the Tx-cooperation phase and prior to ``fast"-decoding phase.)} 
 
 The idea is that in phase $i$ only Txs in $\Tfi:=\mathcal{K}_i \cap \Tf$ send a ``fast" message, all others do not. 
 
 \subsubsection{Transmitting ``fast'' messages in the $i$-th phase} Each active Tx~$k\in\Tfi$  sends its entire ``fast'' message $M_k^{(F)}$ and encodes it  using a non-precoded codeword ${U}_k^{(n)}(M_k^{(F)})$ from a Gaussian codebook of power $\P$. Moreover, during the first Tx-cooperation round, it receives from its two neighbours, Txs $k-1$ and $k+1$, quantized versions of their transmit signals, where quantizations are performed at noise levels. Notice that the neighbouring Txs can share this information because they only send ``slow" messages but no ``fast" messages as they are not in $\mathcal{K}_i$ and thus neither in $\Tfi$.

 Tx~$k\in\Tfi$  computes its input sequence $X_k^n$ as 
 \begin{equation} \label{eq:precoding}
 X_{k}^n = U_k^{n} \big( \MkF\big) - \sum_{\tilde k \in \mathcal I_k^{(S)} } h_{k,k}^{-1} h_{\tilde k, k} \hat X^n_{\tilde k},
 \end{equation}
 where $X^n_{\tilde k}$ denotes the quantized signal of Tx~$\tilde{k}$ and 
 \begin{equation}
 \mathcal I_k^{(S)}  = \{k-1, k+1\} \cap  (\Ta \backslash \Tfi)
 \end{equation}
The  precoding in \eqref{eq:precoding} makes that a ``fast" Rx~$k$ observes the almost interference-free signal
 \begin{equation} \label{eq:14}
 Y_{k}^n = h_{k,k}  U_k^n + \underbrace{\sum_{ \tilde k \in \mathcal I_{k}^{(S)}} h_{ \tilde k ,k} ( X_{ \tilde k}^n - \hat{ X}_{ \tilde k}^n) + Z_{k}^n}_{\textnormal{disturbance}},
 \end{equation}
 where  the variance of above disturbance is around noise level and does not grow with $\P$. Each Rx $k\in\Tfi$  decodes its desired ``fast" message $M_k^{(F)}$ based on \eqref{eq:14}, and during the first Rx-cooperation round it sends the decoded message to their two neighbouring Rxs $k-1$ and $k+1$ so that they can mitigate the interference from ``fast" transmissions.

 \subsubsection{Transmitting ``slow'' messages in the $i$-th phase} 
 We first introduce some notation. Let $k_1, k_2, \ldots$ be the  indices in increasing order of users $k$ for  which $A_k=0$, i.e., of deactivated users. The Tx-Rx pairs lying in between any of these two indices form an independent subnet that does not interfere with the other subnets. We define the users in the j-th subnet as $\mathcal{K}_{\textnormal{subnet},j}:=\{k_{j-1}+1, \ldots, k_{j}-1\}$, where we set $k_{0}=0$, and denoting the random total number of subnets by $J$ we set $k_{J+1}=K+1$.

 We explain the encoding and decoding of ``slow" messages independently for each subnet $j\in\{1,\ldots, J\}$. Let $L_j:= |\mathcal{K}_{\textnormal{subnet},j}|=k_{j}-k_{j-1}-1$ denote the size of this subnet. 
We  split the subnet into smaller non-interfering  subnets of at most $\D+1$ users. Specifically, if  $k_{j-1}+1 \in \Tfi$ or if $\Tfi\cap\mathcal{K}_{\textnormal{subnet},j} =\emptyset$, i.e.,  when the subnet's first transmitter sends a ``fast" message or all Txs in the subnet send ``slow" messages, we  silence all Txs  $k\in\{k_{j-1}+c(\D + 2)\}_{c=1} ^{\lfloor \frac{L_j}{\D+2}\rfloor}$. Otherwise,  we silence all Txs  $k\in\Big\{k_{j-1}+(\D+1), k_{j-1}+(\D+1)+c(\D + 2)\Big\}_{c=1}^{\left\lfloor \frac{L_j-\D-1}{\D+2}\right\rfloor}$. 

In each resulting smaller subnet we apply the following scheme.
%
The first and  last Tx/Rx pairs  in the small subnet apply the coding scheme described above  for ``fast" messages: if the indices of these pairs lie in $\Tfi$, then they send their ``fast" message using this scheme, and otherwise they send parts of their ``slow" message, but using the same scheme. All other ``slow" Tx/Rx pairs of the small subnet apply the CoMP reception scheme as for subnets with only ``slow" transmissions. Here, the Rxs however first precancel the interference from ``fast" transmissions from their receive signals. (Recall that ``fast" Rxs shared their decoded messages during the first Rx-cooperation round with their neighbours.) 
 An example of our scheme is illustrated in Figure~\ref{fig-exp1} for $\D = 6$.

 \subsubsection{MG analysis}
 The described scheme achieves a ``fast'' rate of $R^{(F)} = \frac{1}{2}\cdot \frac{1}{2} \log (1 + \P)$, because each Tx can send its ``fast" message only during one of the two phases, but this message can be decoded based on a interference-free channel. Thus, by \eqref{eq:sf}, the scheme achieves a ``fast'' MG of 
 \begin{equation}\label{eq:SF}
 \S^{(F)} = \frac{\rho \rho_f}{2}.
 \end{equation}

\hn{
 To obtain the average ``slow'' MG, we first calculate the sum MG achieved by this scheme.  To this end,  we notice that \mw{given that} the $j$-th subnet exists \mw{(i.e., $J> j$)} and starts at index $k_{j-1}+1$, then the subnet's random length $L_j$ satisfies: 
  \begin{equation} \label{eq:plk}
 \Pr[L_j =\ell ]= P_{\ell, k_{j-1}+1},  {\mw{\qquad \ell=0,1,2,\ldots,}}
  \end{equation} 
 where for each $k \in \Ta$:
 \begin{equation}\label{eq:sumll}
 P_{\ell,k}:= \begin{cases} \rho^{\ell} (1-\rho) , &\text{if} \;   \ell < K- k+1,\\
 \rho^{\ell}, & \text{if} \; \ell = K- k+1,\\
 0 ,& \textnormal{otherwise}.\end{cases}
 \end{equation} 

 Moreover, in the presented scheme, all scheduled transmissions can be performed at MG $1/2$, because interference can perfectly be mitigated and because we have two equally-long phases $i=1,2$. Therefore, conditioned on the facts that the $j$-th subnet exists, starts at index $k_{j-1}+1$,  and is of length $L_j=\ell$, the random sum-MG achieved over this subnet during phase $i$ is:
 \begin{equation}
 \sumki= \begin{cases}
 \ell - \left \lfloor\frac{\ell}{\D+2}  \right \rfloor ,&\hspace{-9mm} \text{if} \; \; (k_{j-1}+1) \in \mathcal K_i   \\
 & \hspace{-10mm} \text{or} \; \displaystyle \Tfi\cap\mathcal{K}_{\textnormal{subnet},j} =\emptyset,  \\
 \ell - 1 - \left \lfloor\frac{\ell-\D-1}{\D+2} \right \rfloor ,& \textnormal{otherwise}.
 \end{cases}
 \end{equation}
\mw{Setting $A_{0}=0$ with probability $1$, the  expected sum MG over the two phases can be expressed as}:
 	\begin{IEEEeqnarray}{rCl}
 		\lefteqn{ \S^{(S)} + \S^{(F)} } \nonumber \\ 
 		&=& \varlimsup_{K\to \infty} \frac{1}{2} \sum_{i =1}^2  \frac{1}{K} \sum_{k=1}^{K}  \sum_{\ell = 1}^{K-k+1}  \Pr[ A_{k-1} = 0] \cdot P_{\ell,k}\cdot  \S_{\textnormal{sum},i}(k,\ell) \nonumber \\ \\
		&=& \mw{ \varlimsup_{K\to \infty} \frac{1}{K}    \sum_{\ell = 1}^{K} P_{\ell,1}} \notag \\ \notag 
 		&&\qquad  \times \Bigg [ \left (\ell - \left \lfloor \frac{\ell}{\D+2} \right \rfloor  \right )  \left (\frac{1}{2}  + \frac{(1- \rho_f)^{\lfloor \frac{\ell}{2} \rfloor }}{2} \right) \\ 
 		&& \hspace{1cm} +  \left (\ell - 1-\left \lfloor \frac{\ell - \D -1}{\D+2} \right \rfloor  \right ) \left (\frac{1-  (1- \rho_f)^{\lfloor \frac{\ell}{2} \rfloor }}{2} \right)\Bigg ] \nonumber   \\
		 & &+  \varlimsup_{K\to \infty} \frac{1}{K} \sum_{k=2}^{K}   \sum_{\ell = 1}^{K-k+1} P_{\ell,k} (1-\rho)  \notag \\ \notag 
 		&& \times \Bigg [ \left (\ell - \left \lfloor \frac{\ell}{\D+2} \right \rfloor  \right )  \left (\frac{1}{2}  + \frac{(1- \rho_f)^{\lfloor \frac{\ell}{2} \rfloor }}{2} \right) \\ 
 		&& \hspace{0.6cm} +  \left (\ell - 1-\left \lfloor \frac{\ell - \D -1}{\D+2} \right \rfloor  \right ) \left (\frac{1-  (1- \rho_f)^{\lfloor \frac{\ell}{2} \rfloor }}{2} \right)\Bigg ]  \notag\\ \\  \notag
 		& = &\varlimsup_{K\to \infty} \frac{1}{2K}  \sum_{\ell = 1}^{K-1}   \rho^{\ell}  \left [ (K - \ell -1)  (1-\rho)^2 + 2  (1-\rho) \right ] \\ \notag
 		&&\hspace{1cm} \times \Bigg [ 2 \ell - \left \lfloor \frac{\ell}{\D+2} \right \rfloor \left (1 + (1-\rho_f)^{\lfloor \frac{\ell}{2} \rfloor} \right ) \\ \notag
 		&& \hspace{1.4cm}- \left ( 1+ \left \lfloor \frac{\ell-\D-1}{\D+2} \right \rfloor \right )  \left (1 - (1-\rho_f)^{\lfloor \frac{\ell}{2} \rfloor} \right )\Bigg ] \\  
 		&& + \varlimsup_{K\to \infty} \frac{ \rho^K}{2K} \cdot  \Bigg [ 2K- 1 -\left \lfloor \frac{K}{\D+2} \right \rfloor -\left \lfloor \frac{K-\D-1}{\D+2} \right \rfloor    \Bigg] \notag \\ \\
 		& = &\varlimsup_{K\to \infty} \frac{1}{2K}  \sum_{\ell = 1}^{K-1}   \rho^{\ell}  \left [ K (1-\rho)^2  \right ] \notag \\ \notag
 		&&\hspace{1cm} \times \Bigg [ 2 \ell - \left \lfloor \frac{\ell}{\D+2} \right \rfloor \left (1 + (1-\rho_f)^{\lfloor \frac{\ell}{2} \rfloor} \right ) \\ \notag
		 		&& \mw{\hspace{1.4cm}- \left ( 1+ \left \lfloor \frac{\ell-\D-1}{\D+2} \right \rfloor \right )  \left (1 - (1-\rho_f)^{\lfloor \frac{\ell}{2} \rfloor} \right )\Bigg ]} \\  \notag
 		&& - \varlimsup_{K\to \infty} \frac{1}{2K}  \sum_{\ell = 1}^{K-1}   \rho^{\ell}  \left [ (\ell + 1)  (1-\rho)^2 - 2  (1-\rho) \right ] \\ \notag
 		&&\hspace{1cm} \times \Bigg [ 2 \ell - \left \lfloor \frac{\ell}{\D+2} \right \rfloor \left (1 + (1-\rho_f)^{\lfloor \frac{\ell}{2} \rfloor} \right ) \\ \notag
 		&& \hspace{1.4cm}- \left ( 1+ \left \lfloor \frac{\ell-\D-1}{\D+2} \right \rfloor \right )  \left (1 - (1-\rho_f)^{\lfloor \frac{\ell}{2} \rfloor} \right )\Bigg ] \\  
 		&& + \varlimsup_{K\to \infty} \frac{ \rho^K}{2K} \cdot  \Bigg [ 2K- 1 -\left \lfloor \frac{K}{\D+2} \right \rfloor -\left \lfloor \frac{K-\D-1}{\D+2} \right \rfloor    \Bigg].\notag \\  \label{eq:last}
 	\end{IEEEeqnarray}
 \mw{For $\rho=1$, only the  last summand in \eqref{eq:last}  is positive and thus 
 \begin{equation}
  \S^{(S)} + \S^{(F)} = \frac{\D+1}{\D+2}.
   \end{equation}
   Then, by \eqref{eq:SF}:}
 \begin{equation}
 \S^{(S)}  =  \frac{\D+1}{\D+2} -  \frac{\rho \rho_f}{2}.
 \end{equation}
\mw{When $\rho \in (0,1)$, only the  first summand in the asymptotic expression \eqref{eq:last} is non-zero. Moreover, notice that $ 1+ \left \lfloor \frac{\ell-\D-1}{\D+2} \right \rfloor=  \left \lfloor \frac{\ell+1}{\D+2} \right \rfloor$ and thus}
  \begin{IEEEeqnarray}{rCl}
\lefteqn{   \S^{(S)}}\notag  \\
 &=& \frac{(1-\rho)^2}{2}\varlimsup_{K \to \infty}  \sum_{\ell = 1}^{K-1} \rho^{\ell} 2\ell  \notag \\ 
&&-  \frac{(1-\rho)^2}{2} \varlimsup_{K \to \infty}  \sum_{\ell = 1}^{K-1}\rho^{\ell} \left \lfloor \frac{\ell}{\D+2} \right \rfloor \left (1 + (1-\rho_f)^{\lfloor \frac{\ell}{2} \rfloor} \right ) \notag \\ 
&&- \frac{(1-\rho)^2}{2} \varlimsup_{K \to \infty}  \sum_{\ell = 1}^{K-1} \rho^{\ell} \left \lfloor \frac{\ell+1}{\D+2} \right \rfloor \left (1 - (1-\rho_f)^{\lfloor \frac{\ell}{2} \rfloor} \right ) \notag \\ 
&&- \frac{\rho \rho_f}{2} \label{eq:3sums}\\
& \stackrel{{(a)}}{=}& \rho - \frac{\rho \rho_f}{2} -  \frac{(1-\rho^2) \rho^{\D+1}}{2(1-\rho^{\D+2})} \notag \\ 
&& -  \frac{(1-\rho)^2  \rho^{\D+1} (1- \rho_f)^{\frac{\D}{2}}}{2 \left(1- \rho^{\D+2} (1-\rho_f)^{\frac{\D}{2}+1}\right)}, \IEEEeqnarraynumspace
 \end{IEEEeqnarray}
 \mw{where step $(a)$ is obtained by  calculating the three sums in \eqref{eq:3sums} as detailed in the following.} To calculate the first sum, we use the following equality
\begin{IEEEeqnarray}{rCl}
\sum_{x=1}^{\infty} x c ^{x} =  \frac{c}{\mw{(1-c)^2} },   \label{eq:sum2} 
\end{IEEEeqnarray} 
 which is valid for all values of $c \in (0,1)$. \mw{This equality implies:}
 \begin{IEEEeqnarray}{rCl}
 \varlimsup_{K \to \infty}  \sum_{\ell = 1}^{K-1} \rho^{\ell} 2\ell = \frac{2 \rho}{\mw{(1-\rho)}^2}.
\end{IEEEeqnarray}
\mw{To  calculate the second sum, we notice:}
\begin{IEEEeqnarray}{rCl}\label{eq:sum22}
 \lefteqn{  \varlimsup_{K \to \infty}  \sum_{\ell = 1}^{K-1}\rho^{\ell} \left \lfloor \frac{\ell}{\D+2} \right \rfloor \left (1 + (1-\rho_f)^{\lfloor \frac{\ell}{2} \rfloor} \right )} \\ \notag 
&=& \varlimsup_{K \to \infty}  \sum_{j= 1}^{\lfloor\frac{K-1}{\D+2} \rfloor} j \sum_{\ell = j(\D+2)}^{(j+1)(\D+2)-1} \rho^{\ell}  \left (1 + (1-\rho_f)^{\lfloor \frac{\ell}{2} \rfloor} \right ) \\ \notag
&=&  \varlimsup_{K \to \infty}  \sum_{j= 1}^{\lfloor\frac{K-1}{\D+2} \rfloor} j \rho^{j(\D+2)}\sum_{\ell = 0}^{\D+1} \rho^{\ell}  \left (1 + (1-\rho_f)^{\lfloor \frac{\ell + j(\D+2)}{2} \rfloor} \right ) \\ \notag
&=&  \varlimsup_{K \to \infty}  \sum_{j= 1}^{\lfloor\frac{K-1}{\D+2} \rfloor} j \rho^{j(\D+2)}  \sum_{\ell =0}^{\D/2} \rho^{2\ell} \left (1 + (1-\rho_f)^{\ell + \frac{j(\D+2)}{2}} \right) \\ \notag
&& +   \varlimsup_{K \to \infty}  \sum_{j= 1}^{\lfloor\frac{K-1}{\D+2} \rfloor} j \rho^{j(\D+2)} \sum_{\ell =0}^{\D/2} \rho^{2\ell+1} \left (1 + (1-\rho_f)^{\ell + \frac{j(\D+2)}{2}} \right) \\ \notag
&=&  \varlimsup_{K \to \infty}  \sum_{j= 1}^{\lfloor\frac{K-1}{\D+2} \rfloor} j \rho^{j(\D+2)} \sum_{\ell =0}^{\D/2} \rho^{2\ell}  (1+\rho)\left (1 + (1-\rho_f)^{\ell + \frac{j(\D+2)}{2}} \right) \\ \notag
&=&(1+\rho) \varlimsup_{K \to \infty}   \sum_{j= 1}^{\lfloor\frac{K-1}{\D+2} \rfloor} j \rho^{j(\D+2)} \sum_{\ell =0}^{\D/2} \rho^{2\ell} \\ \notag 
&& + (1+\rho) \varlimsup_{K \to \infty}  \sum_{j= 1}^{\lfloor\frac{K-1}{\D+2} \rfloor} j (\rho (1-\rho_f)^{0.5})^{j(\D+2)} \sum_{\ell =0}^{\D/2} (\rho^2(1-\rho_f))^{\ell} \\ \notag 
& \stackrel{{(b)}}{=}& \frac{(1+\rho) \rho^{\D+2}}{(1-\rho^2)(1-\rho^{\D+2})} \\ \notag 
&& + \frac{ (1+\rho) \rho^{\D+2} (1- \rho_f)^{\frac{\D}{2}+1}}{ \left (1-\left ((1-\rho_f) \rho^{2} \right) \right) \left(1- \rho^{\D+2} (1-\rho_f)^{\frac{\D}{2}+1}\right)}.
\end{IEEEeqnarray}
\mw{Here, in step $(b)$, we used   \eqref{eq:sum2} and  the following equality, which holds  for an arbitrary $c$:}
\begin{IEEEeqnarray}{rCl} \label{eq:equalities2}
\sum_{x = 0}^n c^x &=& \frac{1-c^{n+1}}{1-c}, 
\end{IEEEeqnarray}
\mw{to simplify the following asymptotic expressions:}
\begin{subequations}
\begin{IEEEeqnarray}{rCl}
\sum_{\ell = 0}^{\D/2} \rho^{2\ell}& = &\frac{1-\rho^{\D+2}}{1-\rho^2}\\ 
 \sum_{\ell = 0}^{\D/2} \left ((1-\rho_f) \rho^{2} \right)^{\ell} &=& \frac{1-\left ((1-\rho_f) \rho^{2} \right)^{(\D/2+1)}}{1-\left ((1-\rho_f) \rho^{2} \right)}\\ 
 \varlimsup_{K \to \infty} \sum_{j=1}^{\lfloor\frac{K-1}{\D+2} \rfloor} j \rho ^{j(\D+2)} &=  &\frac{\rho^{\D+2}}{(\rho^{\D+2}-1)^2 }  \\
\lefteqn{ \varlimsup_{K \to \infty} \sum_{j=1}^{\lfloor\frac{K-1}{\D+2} \rfloor} j (\rho (1-\rho_f)^{0.5})^{j(\D+2)}}\notag \hspace{3cm}\\ 
& = & \frac{\rho^{\D+2} (1-\rho_f)^{\frac{\D}{2}+1}}{\left (\rho^{\D+2} (1-\rho_f)^{\frac{\D}{2}+1}-1\right)^2 }. \IEEEeqnarraynumspace
\end{IEEEeqnarray}
\end{subequations}
\mw{Finally,  the third sum is simplified in a similar way as the second sum:}
\begin{IEEEeqnarray}{rCl}\label{eq:sum33}
\lefteqn{ \varlimsup_{K \to \infty}  \sum_{\ell = 1}^{K-1}\rho^{\ell} \left \lfloor \frac{\ell + 1}{\D+2} \right \rfloor  \left (1 - (1-\rho_f)^{\lfloor \frac{\ell}{2} \rfloor} \right ) }\\ \notag
&=& \varlimsup_{K \to \infty}  \sum_{j= 1}^{\lfloor\frac{K}{\D+2} \rfloor} j \sum_{\ell = j(\D+2)-1}^{(j+1)(\D+2)-2} \rho^{\ell}  \left (1 - (1-\rho_f)^{\lfloor \frac{\ell}{2} \rfloor} \right ) \\ \notag
&=& \varlimsup_{K \to \infty}  \sum_{j= 1}^{\lfloor\frac{K}{\D+2} \rfloor} j \rho^{j(\D+2)-1}\sum_{\ell = 0}^{\D+1} \rho^{\ell}  \left (1 - (1-\rho_f)^{\lfloor \frac{\ell + j(\D+2)-1}{2} \rfloor} \right ) \\ \notag
&=& \frac{1}{\rho} \varlimsup_{K \to \infty}  \sum_{j= 1}^{\lfloor\frac{K}{\D+2} \rfloor} j \rho^{j(\D+2)}  \sum_{\ell =0}^{\D/2} \rho^{2\ell} \left (1 - (1-\rho_f)^{\ell + \frac{j(\D+2)}{2} -1} \right) \\ \notag
&& + \frac{1}{\rho} \varlimsup_{K \to \infty}  \sum_{j= 1}^{\lfloor\frac{K}{\D+2} \rfloor} j \rho^{j(\D+2)} \sum_{\ell =0}^{\D/2} \rho^{2\ell+1} \left (1 - (1-\rho_f)^{\ell + \frac{j(\D+2)}{2}} \right) \\ \notag
&=& \frac{ (1+\rho)}{\rho} \lim_{K \to \infty}  \sum_{j= 1}^{\lfloor\frac{K}{\D+2} \rfloor} j \rho^{j(\D+2)} \sum_{\ell =0}^{\D/2} \rho^{2\ell} \\\notag 
&& -\frac{1+ \rho(1-\rho_f)}{\rho(1-\rho_f)}\sum_{\ell =0}^{\D/2} \rho^{2\ell}  (1-\rho_f)^{\ell}  \\ \notag 
&& \times \lim_{K \to \infty}  \sum_{j= 1}^{\lfloor\frac{K}{\D+2} \rfloor} j \left (\rho ( 1- \rho_f)^{0.5}\right )^{j(\D+2)}  \\\notag 
& = & \frac{ \rho^{\D+2}}{ \rho(1-\rho) (1-\rho^{\D+2})} \\ \notag 
&&  - \frac{ (1+\rho(1-\rho_f)) \rho^{\D+2} (1- \rho_f)^{\frac{\D}{2}+1}}{\rho (1-\rho_f)  \left (1-\left ((1-\rho_f) \rho^{2} \right) \right) \left(1- \rho^{\D+2} (1-\rho_f)^{\frac{\D}{2}+1}\right)}. 
\end{IEEEeqnarray}
}

\subsection{Scheme 2: Transmitting at $\S^{(F)} = 0$:} 
Similar to scheme 1, except that in each subnet Txs only send ``slow" messages. There is  no need to have two phases and in each subnet $j$ we silence Txs  $k\in\{k_{j-1}+c(\D + 2)\}_{c=1} ^{\lfloor \frac{L_j}{\D+2}\rfloor}$. 

 \subsubsection{MG analysis}  The scheme achieves ``fast" MG
  \begin{equation}
  \S^{(F)}=0.
  \end{equation}
  \hn{We can obtain the average ``slow'' MG \mw{similarly} to before, but now the sum-MG over the $j$-th subnet which starts \mw{at  index $k_{j-1}+1$} and is of length $L_j=\ell$  is:
 \begin{equation} \label{eq:sumll2}
 \S_{\text{sum}} (\mw{k_{j-1}+1}, \ell)=  \ell - \left \lfloor\frac{\ell}{\D+2}  \right \rfloor.
 \end{equation}

 The average ``slow'' MG achieved by this scheme is thus
 \begin{subequations}
  \begin{IEEEeqnarray}{rCl} \label{eq:sonlys}
  \lefteqn{ \S^{(S)}} \notag \\ 
 &=&\varlimsup_{K\to \infty} \frac{1}{K}  \sum_{k=1}^{K}  \sum_{\ell = 1}^{K-k+1} P_{\ell,k}  \cdot \Pr[ A_{k-1} = 0] \cdot \S_{\textnormal{sum}}(k,\ell) \\
 &=&\varlimsup_{K\to \infty} \frac{1}{K} \sum_{k = 1}^K  \sum_{\ell =1}^{K-k+1} P_{\ell, k} \cdot  (1-\rho)  \left (\ell - \left \lfloor \frac{\ell}{\D+2}\right \rfloor \right) \\ 
&=&\varlimsup_{K\to \infty} \frac{1}{K} \sum_{\ell = 1}^{K-1} K  \rho^{\ell} (1 -\rho)^2  \left (\ell - \left \lfloor \frac{\ell}{\D+2}\right \rfloor \right) \\ \label{eq:T2}
&& \hspace{-0.2cm}+\varlimsup_{K\to \infty} \frac{1}{K} \sum_{\ell = 1}^{K-1} \rho^{\ell}  \left ( 2(1 -\rho) - (\ell + 1) \right) \left (\ell - \left \lfloor \frac{\ell}{\D+2}\right \rfloor \right) \nonumber \\ \\ \label{eq:T3}
&& +\varlimsup_{K\to \infty}\frac{\rho^K}{K} \cdot \left (K- \left \lfloor \frac{K}{\D+2}\right \rfloor \right).
 \end{IEEEeqnarray} 
 \end{subequations}
 When $\rho = 1$, then the  average ``slow'' MG is equal to
  \begin{equation}
   \S^{(S)} = 1- \varlimsup_{K\to \infty} \frac{\left \lfloor \frac{K}{\D+2}\right \rfloor}{K} = \frac{\D+1}{\D+2}. 
 \end{equation}
  
When $\rho \in (0,1)$, one can prove that the terms in \eqref{eq:T2} and \eqref{eq:T3} go to zero as   $K \to \infty$, establishing achievability of  ``slow'' MG 
\begin{IEEEeqnarray}{rCl}
\S^{(S)} =  \rho - \frac{(1-\rho)\rho^{\D+2}}{1- \rho^{\D+2}}.
\end{IEEEeqnarray}

\section{Proof of Theorem \ref{theorem2}} \label{sec:converse}
Recall the definitions of $\S_{\text{sum}} (k,\ell) $ and $P_{\ell, k}$ in \eqref{eq:sumll2} and \eqref{eq:plk}. Then for a fixed $K$, \mw{definining again $A_0=0$ with probability~$1$:}
 \begin{subequations}
  \begin{IEEEeqnarray}{rCl} \label{eq:sonlys}
\lefteqn{ \frac{1}{K} \sum_k \left ( \S^{(S)}_k + \S^{(F)} \right) } \\%
&=&\frac{1}{K}  \sum_{k=1}^{K}  \sum_{\ell = 1}^{K-k+1} P_{\ell,k}  \cdot \Pr[ A_{k-1} = 0] \cdot \S_{\textnormal{sum}}(k,\ell) \\ 
 &\stackrel{{(a)}} \le&\mw{ \frac{1}{K}   \sum_{\ell =1}^{K} P_{\ell, 1} \cdot  \left (\ell - \left \lfloor \frac{\ell}{\D+2}\right \rfloor \right)} \nonumber\\ 
 && +  \frac{1}{K} \sum_{k = 2}^K  \sum_{\ell =1}^{K-k+1} P_{\ell, k} \cdot  \left (\ell - \left \lfloor \frac{\ell}{\D+2}\right \rfloor \right) (1-\rho) \\ 
&=& \frac{1}{K} \sum_{\ell = 1}^{K-1} K  \rho^{\ell} (1 -\rho)^2  \left (\ell - \left \lfloor \frac{\ell}{\D+2}\right \rfloor \right) \\ 
&& \hspace{-0.2cm}+ \frac{1}{K} \sum_{\ell = 1}^{K-1} \rho^{\ell}  \left ( 2(1 -\rho) - (\ell + 1) \right) \left (\ell - \left \lfloor \frac{\ell}{\D+2}\right \rfloor \right)  \IEEEeqnarraynumspace\\ 
&& +\frac{\rho^K}{K} \cdot \left (K- \left \lfloor \frac{K}{\D+2}\right \rfloor \right),
 \end{IEEEeqnarray} 
 \end{subequations}
 where the inequality in $(a)$ comes from the fact that the maximum sum MG in a subnet of $\ell$ consecutive active Txs is equal to $\ell - \left \lfloor \frac{\ell}{\D+2}\right \rfloor$. For more details,  see \cite[Chapter 5, Proposition 2] {HomaThesis}.
Letting $K \to \infty$ \mw{proves \eqref{eq:rhoequal1} for $\rho=1$ and  \eqref{eq:conv2} for $\rho \in(0,1)$.}

}

\section{Conclusions and Outlook}
We proposed coding schemes to simultaneously transmit delay-sensitive and delay-tolerant traffic over Wyner's symmetric network with randomly activated users.  In our scheme, each active transmitter always has a ``slow'' (delay-tolerant) data to send and with a certain probability also sends an additional ``fast'' (delay-sensitive) data.  Active transmitters and receivers are allowed to cooperate during total $\D$ rounds  but only ``slow'' transmissions can benefit from cooperation.  We derived inner and outer bound on the MG region. When $\D\to \infty$ or when all the transmitters are active, the bounds coincide and the results show that transmitting ``fast'' messages does not cause any penalty on the sum MG. For finite $\D$ our bounds are still close and prove that the  penalty caused by ``fast'' transmissions is small. This should in particular be considered in view of scheduling algorithms \cite{Bairagi} where transmission of ``fast" messages inherently causes a penalty on the sum-MG that is linear in the ``fast" MG. 

Future interesting research directions include the two-dimensional hexagonal model, which we  studied in \cite{HomaITW2020}. We conjecture that  also for this hexagonal model,  a combination of  Tx- and Rx-cooperation allows to mitigate most of the interference  and  essentially eliminate any penalty caused by transmission of ``fast" messages. As we showed in our previous work \cite{HomaITW2020}, this is not possible  under Rx-cooperation only. Excellent interference cancellation performance is also expected for  multi-antenna setups.   

\section*{Acknowledgment}
The works of M. Wigger and S. Shamai have been supported by the
       European Union's Horizon 2020 Research And Innovation Programme,
       grant agreements no. 715111 for M. Wigger and no. 694630 for S. Shamai. The work of H. Nikbakht and JM Gorce have been supported by the Nokia Bell Labs - Inria common lab, grant agreement ``Network Information Theory''.
  
 \end{document}